\documentclass[aps,prl,twocolumn,superscriptaddress,nopacs,final,nofootinbib,floatfix]{revtex4}

\usepackage{amsmath,amsfonts,amsthm,amssymb}
\usepackage{graphicx}
\usepackage{braket}
\usepackage{setspace}
\usepackage{bbm}
\usepackage{hyperref}
\DeclareMathOperator{\tr}{tr}
\newcommand{\ie}{\textit{i.e.}}
\newcommand{\eg}{\textit{e.g.}}

\begin{document}

\title{Steering bound entangled states: 
A counterexample to the stronger Peres conjecture}

\author{Tobias Moroder}
\affiliation{Naturwissenschaftlich-Technische Fakult\"at, 
Universit\"at Siegen, Walter-Flex-Str.~3, 57068 Siegen, Germany}

\author{Oleg Gittsovich}
\affiliation{Institute of Atomic and Subatomic Physics, TU Wien, 
Stadionallee 2, 1020 Wien, Austria}
\affiliation{Institute for Theoretical Physics, University of Innsbruck, 
Technikerstr. 25, 6020 Innsbruck, Austria}
\affiliation{Institute for Quantum Optics and Quantum Information, 
Austrian Academy of Sciences, Technikerstr. 21a, 6020 Innsbruck, Austria}

\author{Marcus Huber}
\affiliation{Departament de F\'isica, Universitat Aut\`onoma de Barcelona, 
08193 Bellaterra, Spain}
\affiliation{ICFO-Institut de Ci\`encies Fot\`oniques, 
Mediterranean Technology Park, 08860 Castelldefels (Barcelona), Spain}

\author{Otfried G\"uhne}
\affiliation{Naturwissenschaftlich-Technische Fakult\"at, 
Universit\"at Siegen, Walter-Flex-Str.~3, 57068 Siegen, Germany}

\begin{abstract}
Quantum correlations are at the heart of many applications  
in quantum information science and, at the same time, they form 
the basis for discussions about genuine quantum effects and their
difference to classical physics. On one hand, entanglement theory 
provides the tools to quantify correlations in information processing 
and many results have been obtained to discriminate useful entanglement, 
which can be distilled to a pure form, from bound entanglement, 
being of limited use in many applications.
On the other hand, for discriminating quantum phenomena from their 
classical counterparts, Schr\"odinger and Bell introduced the notions 
of steering and  local hidden variable models. We provide a method to 
generate systematically bound entangled quantum states which can 
still be used for steering and therefore to rule out local hidden 
state models. This sheds light on the relations between the various 
views on quantum correlations and disproves a widespread conjecture 
known as the stronger Peres conjecture. For practical applications, 
it implies that even the weakest form of entanglement can be certified 
in a semi-device independent way.
\end{abstract}

\maketitle

Entanglement denotes quantum correlations which cannot be generated 
in any local way. While the characterization of entanglement for pure
two-particle states is straightforward, the task becomes challenging for
noisy or mixed quantum states. Here, even the simple question whether 
or not a given quantum state is entangled is not easy to decide. Apart
from that, it is also difficult to characterize the usefulness of entanglement
for the mixed state case. Since many quantum information protocols like 
quantum teleportation or quantum key distribution work with pure maximally 
entangled states, one may first distill a mixed noisy state to a pure highly
entangled state, but characterizing all possible distillation protocols is
not straightforward. In fact, it was already shown in 1998 that there are 
so-called bound entangled quantum states from which no pure state entanglement
can be distilled \cite{horodecki98a}. This shows some irreversibility in entanglement theory, as these states require pure state entanglement for their generation, but then this entanglement can never be recovered again.

In the following years it turned out that bound entangled states are central to many problems in quantum theory. For instance, it has been shown that entangled states with a positive partial transpose (PPT) are bound entangled, but the question whether all bound entangled states are PPT is, despite of numerous efforts \cite{Dur2000,Shor2001,Watrous2004}, undecided. Using bound entangled states, it has been shown that bound information, an analogue to bound entanglement in classical information theory, exists in the multipartite 
scenario~\cite{Gisin2000}. Furthermore, bound entangled states are conjectured to have a small dimensionality of entanglement~\cite{Sanpera2001}. Finally, it has surprisingly been shown that the correlations of bound entangled states can be used for distilling a secure quantum key~\cite{PhysRevLett.94.160502,horodecki08a}, although no pure state entanglement can be distilled from the state. All these problems and observations clearly justify to call bound entanglement a ``mysterious invention of nature''~\cite{mystery}.

\begin{figure}[t!!]
\includegraphics[width=0.98\linewidth]{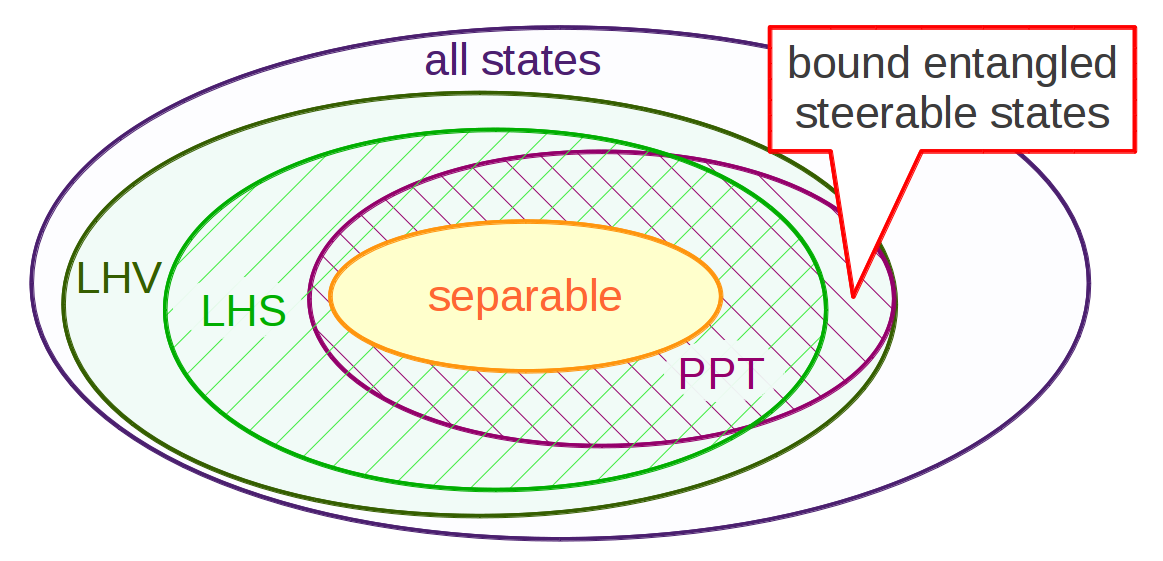}
\caption{A schematic view on the space of all quantum states.
The set of all states is convex with the separable states as 
a subset, states which are not separable are entangled. 
The PPT states are bound entangled, as no pure state 
entanglement can be distilled from these weakly entangled 
states. Some states admit a local hidden state (LHS) model, 
and if this is not the case, they can be used for steering. 
A larger set of states admits a local hidden variable (LHV) 
model, and if this is not the case, the state violates some 
Bell inequality. In this paper we present a method to generate 
PPT states which are steerable. In this figure we have, according
to the Peres conjecture, depicted the PPT states as a subset of 
the LHV states, but the family of states presented in this paper 
may also be outside of the LHV states.}
\label{fig:scheme}
\end{figure}

Besides all the applications in information processing, quantum correlations
are also important when contradictions between quantum mechanics and the classical
world view should be derived. This was highlighted by John Bell, when he showed that 
no local hidden variable model can reproduce the quantum mechanical correlations
\cite{bell64a,brunner_review}. Interestingly, a similar question was discussed
before by Erwin Schr\"odinger, who asked whether one party (called Alice)  
can steer the state from the other party (called Bob) by appropriate measurements,
a task which is not conceivable in a classical world \cite{schroedinger35a,wiseman07a}.
Mathematically, this problem reduces to finding a local hidden {\it state} model 
for the correlations, which is a hidden variable model with the additional constraint
that Bob's measurements are described by the rules of quantum mechanics.

Not surprisingly, bound entanglement is also central to several open problems 
concerning Bell inequalities and steering, see Fig.~\ref{fig:scheme}. Most
prominently, a conjecture by 
Asher Peres~\footnote{Since the conjecture was suggested by Tal Mor, see Ref.~\cite{peres99a}, one could even call it Peres-Mor conjecture.} states that bound (and therefore especially PPT) entangled states 
always admit a local hidden variable model~\cite{peres99a}. It is known that 
this conjecture is wrong in the multipartite case under various different notions of bound entanglement~\cite{duer01a,vertesi12a}, 
but it is still open in the bipartite case. Here it is known to hold true for 
various cases~\cite{werner00a,acin01b,acin02b,masanes06a,salles08a,salles10a,
moroder13a}, but it has also been shown that with the help of additional states and operations, any entangled state shows nonlocal behaviour \cite{masanes08a,Buscemi2012}. Similarly, it has been conjectured that all bound entangled states do admit even a hidden state model and are thus useless for steering scenarios~\cite{pusey13a,skrzypczyk13a}. This conjecture is termed the stronger Peres conjecture and recently strong evidence in favor of it has been claimed~\cite{skrzypczyk13a}. In this paper we disprove it by giving an explicit counterexample. 

More precisely, we present a method to generate systematically bound entangled
states which violate a steering inequality and thus do not admit a hidden state model. This does not only deliver the desired counterexample, it also provides candidates for the other conjectures concerning bound entanglement. For instance, these states are natural candidates for testing the original Peres conjecture or the existence of bipartite bound information \cite{Gisin2000}. Finally, the resulting states are interesting from a practical point of view as their entanglement may be verified in experiments without any assumptions on the measurements on one party. 


\textit{Framework and notation.} Steering can be viewed as
entanglement verification in a nowadays called semi-device independent scenario~\cite{wiseman07a}. One of the parties, let's say Alice, is 
totally untrusted and only the number of settings and respective outcomes 
is specified, while for the other party, Bob, one has a perfect quantum 
description of the measurements.

We consider the case that Alice can choose between different measurements,
each having the same number of possible results. We use $x=1,\dots, m$ to 
label the setting, $a=1,\dots,n$ for the result of the measurement, and 
$a|x$ for the combination. For Bob we assume that he performs full tomography 
on his $d$-dimensional system, so that he can reconstruct the state for each 
possibility $a|x$ of Alice. Thus the available data of this scenario are fully
specified by the ensemble of conditional states for Bob that we describe by 
the collection of unnormalized density operators $\mathcal{E}=\{ \rho_{a|x} \}_{a,x}$ such that $P(a|x)=\tr(\rho_{a|x})$. Note that non-signalling means that 
$\sum_{a} \rho_{a|x}=\rho$ is independent of the setting $x$, and if this 
is fulfilled then the ensemble $\mathcal{E}$ indeed has a quantum representation~\cite{schroedinger35a,hughston93a}. 

Note that in a general steering scenario Bob only measures a few characterized
observables, \eg, only the Pauli matrices $\sigma_x$ and $\sigma_z$, or, 
similarly to Alice, he chooses a setting $y$ and obtains a result $b$ by 
doing a fixed  measurement described by the positive operator valued measure 
$\{M_{b|y}\}_b$. Then the available data are given by the joint conditional 
probability distributions $P(a,b|x,y)$, which admit a local hidden state 
model if they can be written as
\begin{equation}
\label{eq:lhs_slow}
P(a,b|x,y)=\sum_{\lambda} P(\lambda) P(a|x,\lambda) \tr(M_{b|y}\sigma_{\lambda}).
\end{equation}
Here, $\lambda$ is a hidden variable, occuring with probability $P(\lambda)$
and $\sigma_{\lambda}$ are valid quantum states. In contrast to this, a 
local hidden variable model would not have such a constraint for Bob's 
conditional distribution $P(b|y,\lambda)$. Note that any distribution, 
as for instance also $P(a|x,\lambda)$ can still always be written as an 
appropriate measurement on a quantum state~\cite{werner89a} --- via 
this one sees that Eq.~\eqref{eq:lhs_slow} can be obtained by measuring 
a separable state. But the important point is that Bob's measurement is fixed. 

However, since we assume that Bob obtains full tomography, his exact measurement procedure does not matter. If he obtains full information for instance via separate settings and respective outcomes, then the set of all operators $\{M_{b|y}\}_{b,y}$ spans the full operator space, so that the conditions given by Eq.~\eqref{eq:lhs_slow} can only be fulfilled if we have already a corresponding equality on the state space level. Thus, an ensemble 
$\mathcal{E}$ has a local hidden state model if
\begin{equation}
\label{eq:lhs_states}
\rho_{a|x}= P(a|x)\frac{\rho_{a|x}}{\tr(\rho_{a|x})} = 
\sum_{\lambda} P(\lambda) P(a|x,\lambda) \sigma_{\lambda}
\end{equation}
holds for all choices $a|x$. If this is not possible the ensemble $\mathcal{E}$ 
is called steerable, referring to the phenomena that Alice can steer the 
decomposition of Bob's reduced state in a non-trivial way.

Before we proceed, note that the problem of Eq.~\eqref{eq:lhs_states} can 
be simplified if one collects all randomness of Alice's measurement into 
$P(\lambda)$ and Bob's states $\sigma_\lambda$. This accounts to consider 
only the finite number of deterministic strategies for Alice that we 
label by $\lambda_{i_1,i_2,\dots, i_m}$ such that the subscripts $i_k$ 
encode the triggered outcome for each setting, \ie,  $P(a|x,\lambda_{i_1,\dots i_m})=\delta_{i_x,a}$. Then the ensemble $\mathcal{E}$ is non-steerable if 
and only if there exists a set of positive semidefinite operators 
$\omega_{i_1i_2 \dots i_{m}}\geq0$ with $i_k=1,\dots,n$ for each $k=1,\dots,m$ 
such that 
\begin{equation}
\label{eq:class_strat}
\rho_{a|x} = \sum_{i_1, \dots i_{m}} \delta_{i_x,a} \:\omega_{i_1i_2\dots i_{m}}
\end{equation}
holds for all possible $a,x$~\cite{pusey13a}.


\textit{All the steering inequalities.} Hence to show that an ensemble 
$\mathcal{E}$ is steerable one must certify that it is not of the form 
given by Eq.~\eqref{eq:class_strat}. This certificate is called steering 
inequality and is similar in spirit to Bell inequalities~\cite{bell64a,clauser69a,peres99a} or entanglement witnesses~\cite{terhal00a}. A linear steering inequality is a linear 
function of the given ensemble $C(\mathcal{E})$ such that  
$C(\mathcal{E})\geq 0$ holds for all non-steerable ensembles 
$\mathcal{E}$, so that  $C(\mathcal{E})<0$ witnesses steering.

In order to derive the form of all such steering inequalities 
one uses Eq.~\eqref{eq:class_strat} and can proceed along the 
following lines: The question given Eq.~\eqref{eq:class_strat} 
is a special convex optimization problem called semidefinite 
programming, \ie, 
$\min_{x\in\mathbbm{R}^n}\{c^Tx|F_0+\sum_i x_i F_i \geq0\}$ 
with $c\in\mathbbm{R}^n$ and Hermitian matrices $F_0$ and all $F_i$. Due to the 
convex structure of the problem one can solve the alternative problem, called 
dual $\max_{Z\geq 0} \{-\tr(ZF_0)|\tr(ZF_i)=c_i \}$, which lower bounds the original problem and finally 
attains the same optimal value. This dual problem is effectively the optimization over all steering inequalities. Thus to derive all steering inequalities we just need to put Eq.~\eqref{eq:class_strat} into the form of a semidefinite program and invoke its dual. This approach has been used in a quantification of steering~\cite{skrzypczyk13a}.

For our main intended goal we consider only a steering inequality for the 
case $m=2$ and $n=3$, since this be the setting of our counterexample. It should be noted, however, that
our approach can directly be applied also for more than two measurements 
or more outcomes. In the following we state the form of all such 
inequalities and verify $C(\mathcal{E})\geq0$ for all non-steerable ensembles. 


\textit{Steering inequality.} 
Consider the described steering scenario for $m=2$ and $n=3$. Suppose we 
have a set of operators $\mathcal{Z}=\{ Z_{13},Z_{23},Z_{31},Z_{32},Z_{33} \}$, 
each positive semidefinite $Z\geq 0 $ for all $Z \in \mathcal{Z}$, and further 
satisfying
\begin{align}
\label{eq:Z1}
Z_{11}&=Z_{13}+Z_{31}-Z_{33}\geq 0, \\
\label{eq:Z1b}
Z_{21}&=Z_{23}+Z_{31}-Z_{33}\geq 0, \\
\label{eq:Z1c}
Z_{12}&=Z_{13}+Z_{32}-Z_{33}\geq 0,\\
\label{eq:Z1d}
Z_{22}&=Z_{23}+Z_{32}-Z_{33}\geq 0.
\end{align}
Then the linear function 
\begin{align}
\nonumber
C(\mathcal{E})\!=\!& \tr(Z_{13} \rho_{1|1})\! +
\! \tr(Z_{23} \rho_{2|1})\! +\! \tr(Z_{31} \rho_{1|2})\! + 
\!\tr(Z_{32} \rho_{2|2} ) \\ \label{eq:steering_inequality}
& + \tr[Z_{33} (\rho - \rho_{1|1} - \rho_{2|1} - \rho_{1|2} - \rho_{2|2})]
\end{align}
is non-negative for all non-steerable ensembles of 
Eq.~\eqref{eq:lhs_states}. Thus $C(\mathcal{E})<0$ shows that 
the ensemble is steerable.

To show this,  note that a given ensemble $\mathcal{E}$ with $m=2$ and $n=3$ is non-steerable if and only if there exists $\omega_{ij}\geq 0$ with $i,j=1,2,3$ such that 
\begin{align}
\rho_{1|1}&=\omega_{11} + \omega_{12} + \omega_{13},& \rho_{1|2}&=\omega_{11}+\omega_{21}+\omega_{31}, \nonumber \\
\rho_{2|1}&=\omega_{21} + \omega_{22} + \omega_{23},&
\rho_{2|2}&=\omega_{12}+\omega_{22}+\omega_{32},
\label{eq:omega1}
\end{align}
and
\begin{equation}
\label{eq:omega2}
\rho= \rho_{1|1} + \rho_{2|1} + \rho_{3|1} = \rho_{1|2} + \rho_{2|2} + \rho_{3|2} = \sum_{ij} \omega_{ij}
\end{equation}
hold. 
Using these relations in Eq.~\eqref{eq:steering_inequality} one can verify 
that this expression equals to $C(\mathcal{E})=\sum_{ij}\tr(Z_{ij}\omega_{ij})$ 
and hence is non-negative since all occuring operators are positive semidefinite.


\textit{Strategy for generating counterexamples.} 
Now we can present our method of generating bound entangled 
states. Let us assume that we have fixed a linear steering 
inequality, \ie, a set of valid operators $\mathcal{Z}$ satisfying 
the conditions from the previous section. From this one can obtain an entanglement witness~\cite{terhal00a} by employing any possible choice of measurements for Alice in the expression $C$. 

For the case of $n=2,m=3$ this means that the operator
\begin{align}
\nonumber
W=& A_{1|1}\otimes Z_{13} + A_{2|1} \otimes Z_{23} + A_{1|2} \otimes Z_{31} + A_{2|2} \otimes Z_{32} \\
\label{eq:witness}
&+ (\mathbbm{1}- A_{1|1} - A_{2|1} -A_{1|2}-A_{2|2} ) \otimes Z_{33}
\end{align}
is non-negative on separable states for any set of operators $A_{a|x}$ 
satisfying $A_{a|x} \geq 0$ and $\sum_a A_{a|x}=\mathbbm{1}$ for all 
combinations $a,x$ and one readily gets $C=\tr(W\rho_{AB})$.

The method is then as follows: We assume that Alice and Bob both 
have qutrits and that Alice makes a projective measurement in two 
mutually unbiased bases. After that we look for a ``good'' steering 
inequality, \ie, a good set $\mathcal{Z}$. To do so we randomly choose 
a pure state, compute its ensemble $\mathcal{E}$ using the fixed 
measurements of Alice, and determine the best steering inequality 
$\mathcal{Z}$. Afterwards we build up the given witness $W$ and 
minimize its expectation value with respect to all PPT states. If 
this optimum is negative, then we have already a counterexample. 
If this fails then we start over. However, once we found a PPT 
state violating the randomly chosen steering inequality, we can 
use this state, compute its ensemble and look for an even better 
steering inequality. And similarly, once we have a better steering
inequality we can look for an even  better state. In this sense we 
further amplify the violation of the PPT entangled state and we 
repeat this procedure until the violation saturates.

Note that the occurring optimizations are semidefinite programs 
and thus can be done efficiently~\cite{yalmip,sedumi}. 
Furthermore, we should add that we normalize the steering 
inequality such that each $Z \in \mathcal{Z}$ satisfies $\tr(Z)=1$.


\textit{Counterexample.} Running the explained procedure quickly 
results in bound entangled states which serve as counterexamples
to the stronger Peres conjecture. Interestingly, if one 
amplifies the violation, we always end up with a maximal 
violation of $C=-0.0029$. From the numerical solution one 
can infer the 
following analytical solution:

At first let us describe the steering inequality: The set of operators $Z_{13}=\ket{q_+}\bra{q_+}$, $Z_{23}=\ket{q_{-}}\bra{q_{-}}$, $Z_{32}=Z_{33}=\ket{s}\bra{s}$  and $Z_{31}=(1-x)\ket{t}\bra{t}+x\ket{2}\bra{2}$ 
with real, normalized vectors
\begin{align}
\ket{q_\pm}&=[a,\sqrt{1-a^2-b^2},\mp b],\nonumber \\
\ket{s}&=[a,-\sqrt{1-a^2},0], \nonumber \\
\ket{t}&=[c,-\sqrt{1-c^2},0]
\end{align} 
and abbreviations
\begin{align}
a=&\sqrt{\frac{2}{3}(1+x)}, \:\:b=\sqrt{\frac{1}{4}(1-2x)}, \nonumber \\
c&=\sqrt{\frac{2}{3}(1-2x)}/\sqrt{1-x}
\end{align}
defines a one-parameter family of steering inequalities 
for $0\leq x\leq 1/2$.

This can be seen as follows: With this ansatz we already fulfil 
the positivity requirements of each individual $Z\in \mathcal{Z}$. 
Moreover, the additional constraints given by Eqs.~(\ref{eq:Z1c}, \ref{eq:Z1d})
are satisfied automatically since $Z_{32}=Z_{33}$, while 
from Eqs.~(\ref{eq:Z1}, \ref{eq:Z1b}) we only need to check one condition,
since the unitary matrix $V=\textit{diag}(1,1,-1)$ 
interchanges $Z_{13}$ with $Z_{23}$, (i.e.,~$Z_{23}=VZ_{13}V^\dag$), but 
leaves $Z_{31}$ and $Z_{33}$ invariant. Thus we only need to show that 
$Z_{11}\geq0$, for which the particular choices of $a,b,c$ become important. 
These are determined by the identity $Z_{13}+Z_{23}+Z_{31}=\textit{diag}(2,1/2,1/2)$ 
that we observed from the numerical solution. Via this choices the operator 
$Z_{11}$ then has the same eigenvalues as $Z_{31}$, \ie, eigenvalues 
$\{x,1-x,0\}$. 

\begin{figure}[t!!]
\includegraphics[width=0.95\linewidth]{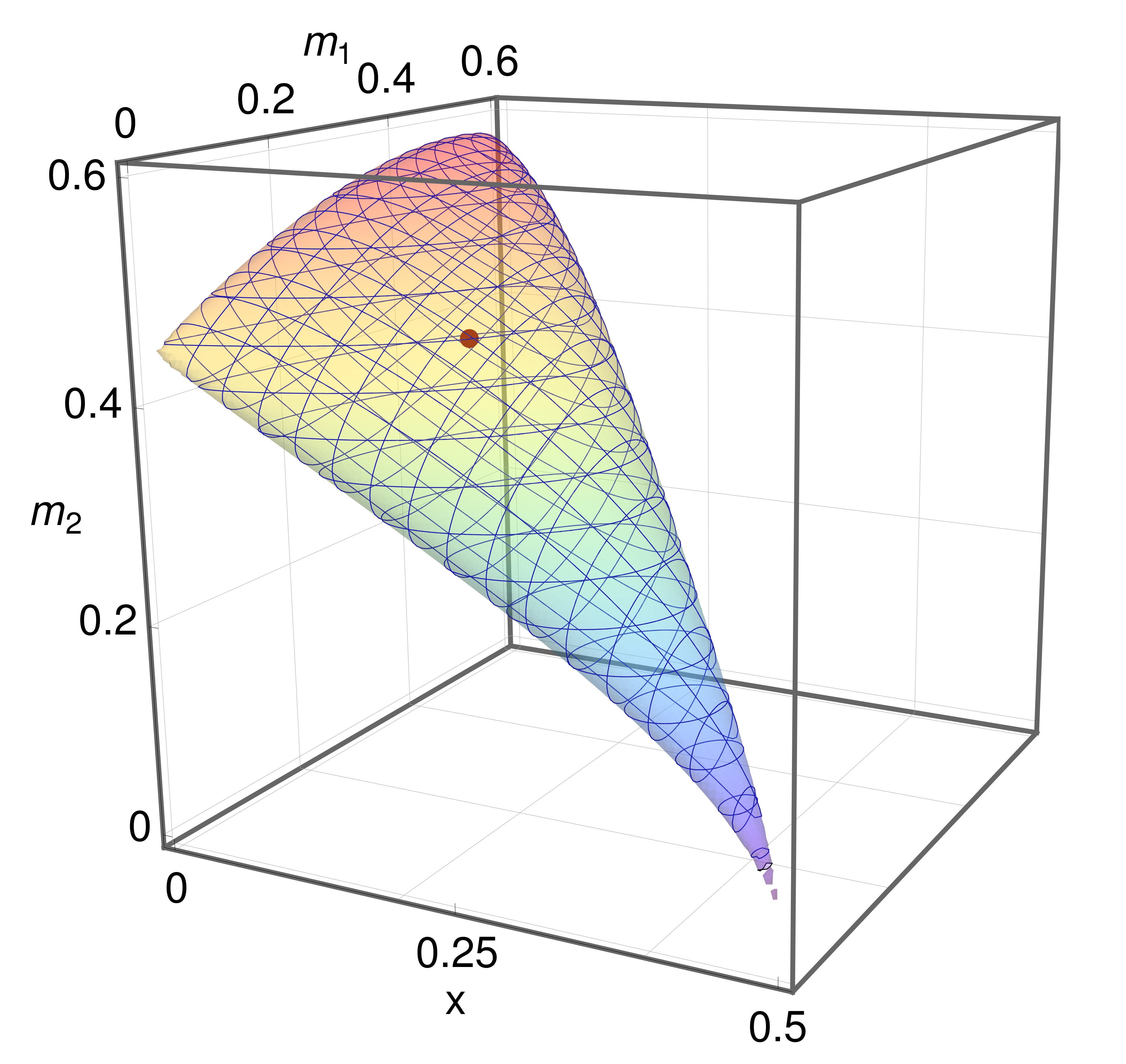}
\caption{The family of states which are counterexamples to
the stronger Peres conjecture. The parameters $m_1$ and $m_2$
characterize the state, while $x$ characterizes the steering
inequality. The red dot corresponds to the values 
$x=0.1578,m_1=0.2162,m_2=0.4363$, resulting in
the example given in Eq.~(\ref{examplerho}) and leading 
to a high violation of the steering inequality.}
\label{fig:violation}
\end{figure}

Second, before discussing the state, let us fix the two mutually 
unbiased bases,  since we employ here already some rotated form 
which makes the final bound entangled state look simpler. 
The respective vectors are denoted by $\ket{v_{x|a}}$ 
and are given by
\begin{align}
\ket{v_{1|1}}&=[1/\sqrt{3},-1/\sqrt{6},-1/\sqrt{2}],\nonumber \\
\ket{v_{2|1}}&=[1/\sqrt{3},-1/\sqrt{6},1/\sqrt{2}],\nonumber \\
\ket{v_{3|1}}&=[1/\sqrt{3},\sqrt{2/3},0],
\end{align}
for the setting $a=1$ and 
\begin{align}
\ket{v_{1|2}}&=[1,0,0],\nonumber \\
\ket{v_{2|2}}&=[0,q/\sqrt{2},\mathbbm{i}q/\sqrt{2}],\nonumber \\
\ket{v_{3|2}}&=[0,q^*/\sqrt{2},-\mathbbm{i}q^*/\sqrt{2}],
\end{align}
with $q=(-1)^{2/3}$ for setting $a=2$. 

Finally, let us turn to the state: Consider the following class of states
\begin{align}
\nonumber
\rho_{AB}&=\lambda_1\ket{\psi_1}\bra{\psi_1}+\lambda_2\ket{\psi_2}\bra{\psi_2} \\ &+\lambda_3 (\ket{\psi_3}\bra{\psi_3}+\ket{\tilde \psi_3}\bra{\tilde \psi_3})
\end{align}
using the normalized states
\begin{eqnarray}
\ket{\psi_1}&=&(\ket{12}+\ket{21})/\sqrt{2}, \nonumber \\
\ket{\psi_2}&=&(\ket{00}+\ket{11}-\ket{22})/\sqrt{3},  \nonumber \\
\ket{\psi_3}&=&m_1\ket{01}+m_2\ket{10}+m_3(\ket{11}+\ket{22}),\nonumber \\
\ket{\tilde \psi_3}&=&m_1\ket{02}-m_2\ket{20}+m_3(\ket{21}-\ket{12}) 
\end{eqnarray}
with $m_i\geq 0$. 

By construction, this represents a valid quantum state. In order 
to assure that this state has a positive partial transpose, we 
make it PPT invariant, \ie, $\rho_{AB}=\rho_{AB}^{T_A}$, for 
which one must make the off-diagonal blocks Hermitian. These 
constraints will fix the eigenvalues to
\begin{align}
\lambda_1&=1-\frac{2+3m_1m_2}{4-2m_1^2+m_1m_2-2m_2^2},\\
\lambda_3&=\frac{1}{4-2m_1^2+m_1m_2-2m_2^2}.
\end{align}
The parameter $\lambda_2=1-\lambda_2-2\lambda_3$ is then given 
by normalization. The $\lambda_i$ are therefore parametrized
by $m_1,m_2$ and this is only giving non-negative eigenvalues if 
$m_1^2+m_2^2+m_1m_2\leq 1$.

Summarizing, we have deduced a class of steering inequalities 
$\mathcal{Z}$ parametrized by $0\leq x\leq 1/2$, a set of 
measurements for Alice given by the two mutually unbiased 
bases, and a class of PPT states that depend on two non-negative, 
constrained parameters $m_1,m_2$. For these choices one can 
now  compute expectation values of the steering inequality, 
and deduce combinations which verify steering, see Fig.~\ref{fig:violation}.

A simple optimization over these parameters gives $C=-0.0029$ 
for  $x=0.1578,m_1=0.2162,m_2=0.4363$. These parameters then 
result in vectors 
\begin{align}
\ket{q_\pm}&=[0.8785,0.2388,\mp0.4137],\nonumber \\
\ket{s}&=[0.8785,-0.4777,0], \nonumber \\
\ket{t}&=[0.7361,-0.6769,0].
\end{align}
which define the steering inequality and the state 
\begin{widetext}
\begin{equation}
\rho_{AB}=\left[
\begin{array}{ccccccccc}
0.026 &0 &0 &0 &0.0261 &0 &0 &0 &-0.0261 \\
0 &0.0129 &0 &0.0261 &0.0369 &0 &0 &0 &0.0369 \\
0 &0 &0.0129 &0 &0 &-0.0369 &-0.0261 &0.0369 &0 \\
0 &0.0261 &0 &0.0526 &0.0744 &0 &0 &0 &0.0744 \\
0.0261 &0.0369 &0 &0.0744 &0.132 &0 &0 &0 &0.0792 \\
0 &0 &-0.0369 &0 &0 &0.29 &0.0744 &0.0792 &0 \\
0 &0 &-0.0261 &0 &0 &0.0744 &0.0526 &-0.0744 &0 \\
0 &0 &0.0369 &0 &0 &0.0792 &-0.0744 &0.29 &0 \\
-0.0261 &0.0369 &0 &0.0744 &0.0792 &0 &0 &0 &0.132
\end{array} \right].
\label{examplerho}
\end{equation}
\end{widetext}

That this state is entangled can for instance be checked directly 
via the  covariance matrix criterion~\cite{guehne1}. Alternatively, 
one can even verify that the state does not possess a symmetric 
extension to two copies of $A$~\cite{doherty04a}, which is a necessary 
requirement that the state does not admit a local hidden state model 
with two settings for $A$~\cite{terhal03a}.
Note that all numerical values are rounded up to the $4$-th digit. 
One could even take these values explicitly to convince oneself 
about the counterexample. If one is worried about the numerical 
precision, then we like to emphasize that all conditions of 
$\mathcal{Z}$ and $\rho_{AB}$ are easy to check. Using the 
given $4$-digit representation as the actual ones, one can 
add for instance a small proportion of the identity to form 
the operators $\tilde Z = (1-\epsilon) Z + \epsilon \mathbbm{1}$ 
and the state $\tilde \rho_{AB}=(1-\epsilon)\rho_{AB}/\tr(\rho_{AB}) + \epsilon \mathbbm{1}/9$ such that all constraints are fulfilled 
and that the smallest eigenvalue is well above the precision.
For instance taking $\epsilon=10^{-3}$ amounts that the smallest
eigenvalue of all conditions is well above $10^{-5}$ while the 
violation is still significant $C=-0.0014$. 


\textit{Conclusion}. We provided a way to generate bound entangled states which do not possess 
a local hidden state model and thus violate a steering inequality. This 
disproves the stronger Peres conjecture and shows that the original 
Peres conjecture cannot be proven by considering the stronger steering 
case. It also means that even the weakest form of entanglement can 
be verified in a semi-device independent way.

Naturally, the generated bound entangled quantum states are interesting 
candidates for some of the conjectures concerning bound entanglement. A
first question is whether with a few further modifications of our states 
and measurements one could even find a violation of a Bell inequality 
and thus disprove also the original Peres conjecture. 
A second question is whether this bound entangled state could even allow the generation of a secret key in a semi-device independent quantum key distribution protocol. Third, these bound entangled states even give prominent candidates to investigate whether they could be useful for teleportation or in entanglement swapping in quantum repeaters~\cite{bauml14a}. Finally, it would be interesting to use our method to generate bound entangled states in higher dimensions, such as a 4$\otimes$4 system which be viewed as a four-qubit system. Given the recent advances in quantum control, such states could probably be observed with entangled photons or ions.

\begin{acknowledgments}
We would like to thank J.-D. Bancal, N. Brunner, M. Navascu\'es 
and Y.~C. Liang for stimulating discussions about the Peres conjecture. 
This work has been supported by the EU (Marie Curie CIG 293993/ENFOQI 
and Marie Curie IEF 302021/QUACOCOS), the BMBF (Chist-Era Project QUASAR), 
the FQXi Fund (Silicon Valley Community Foundation), the DFG, the 
Austrian Science Fund (FWF) and the Marie Curie Actions 
(Erwin Schr\"odinger Stipendium J3312-N27).
\end{acknowledgments}

{\it Note added.} After appearance of our results on the 
arXiv it was noted that if one takes the state from our family
with $m_1 = 1/60$ and $m_2=3/10$, the MUB measurements of Alice and the three dichotomic measurements of Bob characterized by the steering inequality, more precisely the $Z_{13}, Z_{23}, Z_{33}$ with $x=0.26$, then the corresponding data do not admit a local hidden variable model~\cite{TamasNicolas}. This shows that our method can indeed be used to find counterexamples to the original Peres conjecture. Note however that this approach is not working for the optimal steering parameters.


\bibliography{bib_dvi}

\end{document}